\pdfoutput=1
\documentclass[12pt]{article} 
\usepackage{geometry}
\usepackage{amsmath} % already in sashorthand but should be defined earlier here to use \numberwithin
\numberwithin{equation}{section}
\RequirePackage[pdftex,dvipsnames]{xcolor}
\definecolor{darkblue}{rgb}{0.1,0.1,.7}
\usepackage[colorlinks, linkcolor=darkblue, citecolor=darkblue, urlcolor=darkblue, linktocpage]{hyperref} 

\usepackage[square, comma, sort&compress,numbers]{natbib} % reference management

\usepackage[margin=10pt,font=small,labelfont=bf]{caption}

\definecolor{brightpink}{rgb}{1.0, 0.0, 0.5}

\usepackage{hyperref}
\hypersetup{colorlinks,linkcolor={brightpink},citecolor={brightpink},urlcolor={black}}  
\usepackage{skt}

\newcommand*\JJskip{8mu}
\catcode`,\active
\newcommand*\JJ{\begingroup
	\catcode`\,\active
	\def ,{\mskip\JJskip\relax}%
	\doJJ
}
\catcode`\,12
\def\doJJ#1#2#3#4{%
	\Phi^{#1}_{#2}\biggl[\genfrac..{0pt}{}{#3}{#4}\biggr]%
	\endgroup
}

\usepackage{tikz}
\usetikzlibrary{external}

\usepackage[]{sashorthand} % Provides useful commands and packages. It has as three potions:
\usepackage[final]{pdfpages}

\DeclareMathAlphabet{\mathpzc}{OT1}{pzc}{m}{it} % New font \mathpzc
\graphicspath{ {images/} }
\begin{document}
	\vspace*{-.6in} \thispagestyle{empty}
	\begin{flushright}
	\end{flushright}
	\vspace{.2in} {\Large
		\begin{center}
			{\bf On spinning loop amplitudes in Anti-de Sitter space\vspace{.1in}}
		\end{center}
	}
	\vspace{.2in}
	\begin{center}
		{\bf 
			Soner Albayrak$^\text{\bf{B}}$ and Savan Kharel$^{\text{\bf{L}}}$$^,$$^\text{\bf{M}}$}
		\\
		\vspace{.2in} 
		$\text{\bf{B}}$ {\it Department of Physics, Yale University, New Haven, CT 06511}\\ 
	         $\text{\bf{L}}$ {\it  Department of Physics, Williams College, MA, 01267} \\
		$\text{\bf{M}}$ {\it  Department of Physics, University of Chicago, Chicago, IL 60637}
	\end{center}
	
	\vspace{.2in}
	
\begin{abstract}
In this work we present a systematic study of AdS$_{d+1}$ loop amplitudes for gluons and gravitons using momentum space techniques. Inspired by the recent progress in tree level computation, we construct a differential operator that can act on a scalar factor in order to generate gluon and graviton loop integrands: this systematizes the computation for any given loop level Witten diagram. We then give a general prescription in this formalism, and discuss it for bubble, triangle, and box diagrams.
\end{abstract}
	
	\newpage
	
	\tableofcontents
	
%	\newpage
	
% BODY
\section{Introduction}

The gauge gravity duality or the AdS/CFT is the correspondence between weakly coupled theories of gravity in Anti-de Sitter space and conformal field theories with large $N$. This correspondence provides a powerful framework to study quantum gravity on Anti-de Sitter space\cite{Maldacena:1997re,Gubser:1998bc,Witten:1998qj}. Given the importance of this duality, a lot of effort has been invested to compute tree level AdS scattering amplitudes in configuration space and Mellin space \cite{Freedman:1998bj,Liu:1998ty,DHoker:1999mqo,Penedones:2010ue,Paulos:2011ie, Fitzpatrick:2011ia, Kharel:2013mka, Fitzpatrick:2011hu,Costa:2014kfa, Hijano:2015zsa, Ghosh:2018bgd,Penedones:2019tng, Caron-Huot:2018kta, Zhou:2020ptb, Alday:2020lbp}.  In the recent years, there has been some renewed interest in computing CFT correlators in momentum space \cite{Raju:2010by,Raju:2012zs, Raju:2012zr, Raju:2011mp, Albayrak:2018tam, Albayrak:2019asr, Albayrak:2019yve, Albayrak:2020isk, Bzowski:2013sza, Bzowski:2015pba,Bzowski:2018fql, Bzowski:2015yxv, Bzowski:2015pba, Bzowski:2019kwd, Isono:2019ihz, Isono:2018rrb,Isono:2019wex, Coriano:2013jba, Coriano:2018bbe, Anand:2019lkt, Gillioz:2019lgs, Farrow:2018yni, Nagaraj:2019zmk, Nagaraj:2020sji, Jain:2020rmw, Oh:2020zvm}.\footnote{There has also been recent results in $p$-adic space \cite{Jepsen:2018ajn, Jepsen:2018dqp, Jepsen:2019svc}. Additionally, because of translation invariance momentum space is a natural choice  for cosmological correlators. For some related recent papers, see \cite{Arkani-Hamed:2017fdk, Arkani-Hamed:2018bjr, Benincasa:2018ssx, Benincasa:2019vqr, Arkani-Hamed:2018kmz,Baumann:2019oyu, Baumann:2020dch, Sleight:2019mgd, Sleight:2019hfp}.}

However, most of the progress is largely focused on tree level results. AdS loop amplitudes pose difficult technical problems.\footnote{It is interesting to note that de Sitter loops are also conceptually difficult. For instance it was pointed out in \cite{Weinberg:2005vy} that scale factor $a(t)$ enters the logarithmic divergence. For some recent progress in de Sitter loops, see \cite{Gorbenko:2019rza}} In addition to the standard loop integrals, one performs bulk integrals whose complexity is already comparable to loop integrals in flat space. For a long time, there were very few loop-level results; however, some progress has occurred in last few years. In \cite{Fitzpatrick:2011hu,Fitzpatrick:2011dm}, Mellin amplitudes corresponding to loop Witten diagrams in AdS were used to study analytical properties of such amplitudes. These papers inspired the usage of CFT crossing symmetry  \cite{Aharony:2016dwx} which lead to progress in computing loops in  AdS$_5$ $\times S^5$ \cite{Alday:2017xua, Aprile:2017bgs, Aprile:2017qoy,Alday:2017vkk}. Progress in the computation of scalar loop diagrams  was performed recently in \cite{Bertan:2018afl,Beccaria:2019stp,Bertan:2018khc, Albayrak:2020isk}. Some progress in studying unitarity in the context of AdS was carried out in \cite{Fitzpatrick:2011dm} and more recently in \cite{Ponomarev:2019ofr, Meltzer:2019pyl, Meltzer:2019nbs}. In \cite{Liu:2018jhs}, it was shown that higher-point diagrams at one-loop may be written in terms of the $6j$ symbols of the conformal group. Similarly, Mellin space pre-amplitudes and the pole structure of the result was investigated in \cite{Yuan:2017vgp,Yuan:2018qva}. In \cite{Carmi:2018qzm, Carmi:2019ocp}, 1-loop bubble diagram in spectral representation for a $\phi^4$ scalar was performed. An algorithm which computes the one-loop Mellin amplitudes for AdS supergravity was demonstrated in \cite{Alday:2019nin}.\footnote{Also, for string theory corrections to such one loop amplitudes, see \cite{Chester:2019pvm, Alday:2020lbp}} Similarly cutkosky rules in CFT's at both strong and weak coupling is studied in \cite{MeltzerAllic}.

Despite the aforementioned progress, work in loop amplitudes is still in a developing stage. It was shown in \cite{Albayrak:2018tam, Albayrak:2019yve, Albayrak:2019asr} that higher point  gravity and gauge theory tree amplitude takes a simplified form  with the judicious use of momentum space formalism. We view our work as the natural extension of tree level results in gauge and gravity theory with the usage of momentum space. We are inspired by the stunning progress in the study of flat space S-matrix at loop-level which has revealed powerful mathematical structures and remarkable physical insight. Many of the results in flat space loop calculations have shown the connection between trees and loops \cite{Bern:1994cg,CaronHuot:2010zt} and gravitational theories to gauge theories \cite{Bern:2008qj}, and the loop amplitudes also correspond to geometric structures \cite{Arkani-Hamed:2013jha}. Many of these deep connections and powerful mathematical structures  have occurred in the context of gauge and gravity theory and with the usage of momentum space. We initiate this investigation as we are interested in exploring whether the AdS loop level gauge and gravity theory scattering amplitudes encodes analogous rich structures to flat space scattering amplitudes.

Here is the organization of the paper. In \secref{\ref{sec: review}}, we review the AdS momentum space formalism on tree-level amplitudes for gauge and gravity theory and discuss the necessary modifications to extend them beyond tree level computations. In particular, we manage to write any loop-level Witten diagram as a differential operator acting on a scalar factor. In \secref{\ref{sec: scalar factors}}, we further discuss these scalar factors by providing implicit results for gluon triangle and box diagrams and by going over the explicit computation of gluon bubble diagram. We then conclude with future directions. Many technical details are collected in appendices.

\section{Momentum space formalism: review of tree level technology and extension to loops}
\label{sec: review}

We start by defining the bulk to boundary propagators\footnote{The polarization vector $\bm{\epsilon}^i_\mu$ also has color dependence but we suppress it and we work with color-ordered gluon diagrams throughout the paper.}
\be 
\label{eq: bulk to boundary propagators in terms of scalar propagator}
t_{\mu}(z,k_i)\equiv\bm{\epsilon}^i_\mu\f^{d-2}_{d-2}(k_i,z)
\;,\qquad
t_{\mu\nu}(z,k_i)\equiv\bm{\epsilon}^i_{\mu\nu}
\f^{d-4}_{d}(k_i,z)
\ee
where $i$ labels different external legs and where we define
\be 
\f^\mu_\nu(k,z)\equiv  \sqrt{\frac{2}{\pi}}z^{\frac{\mu}{2}} k^{\frac{\nu}{2}} K_{\frac{\nu}{2}}(k z)
\ee  
for convenience. We note that all propagators in this paper are in axial gauge, similar to our previous work \cite{Albayrak:2018tam, Albayrak:2019asr, Albayrak:2019yve}. The bulk to bulk propagators read as
\be
\label{eq: traditional propagators}
\mathcal{G}_{\mu\nu}(\bm{k}; z, z')=&
\Pi^{(1)\bm{k}}_{\mu\nu}\int\limits_0^\infty p dp \;
\JJ{d-2}{d-2}{k,p}{z,z'}
+
\Pi^{(2)\bm{k}}_{\mu\nu}
\int\limits_0^\infty
pdp
\frac{k^2+p^2}{p^2} 
\JJ{d-2}{d-2}{k,p}{z,z'}
\\
\mathcal{G}_{\mu\nu\rho\sigma}(\bm{k}; z, z')=&
\Pi_{\mu\nu\rho\sigma}^{(1)\bm{k}}\int
p dp
\JJ{d-4}{d}{k,p}{z,z'}+
\Pi_{\mu\nu\rho\sigma}^{(2)\bm{k}}\int
p dp\frac{k^2+p^2}{p^2}
\JJ{d-4}{d}{k,p}{z,z'}\\
&+
\Pi_{\mu\nu\rho\sigma}^{(3)\bm{k}}\int
p dp\frac{k^2\left(k^2+p^2\right)}{p^4}
\JJ{d-4}{d}{k,p}{z,z'}
\ee
where we define the shorthand notation
\be 
\JJ{\mu}{\nu}{k,p}{z,z'}\equiv \frac{\left(z z'\right)^{\frac{\mu}{2}} J_\frac{\nu}{2}(p z)J_\frac{\nu}{2} (p z')}{k^2+p^2-i\epsilon}
\ee 
for brevity and where $\Pi$ are projectors  that depend on the vector $\bm{k}_\mu$ and the boundary metric $\eta_{\mu\nu}$:  we refer the reader to Appendix~\ref{sec: projectors and differential operators} for the explicit form of any object without definition in this section. We also note that we are working in the Poincar\'e patch of the AdS with the metric \mbox{$ds^2=z^{-2}\left(dz^2+\eta_{\mu\nu}dx^\mu dx^\nu\right)$}.

The relevant three and four point vertex factors for gluons  and three point vertex factor for graviton are as follows\footnote{The overall $z^{4,8}$ factors follow from the inverse metrics that needed to be contracted with to write $V$ in contravariant form.}
\be
V^{\mu\nu\rho}_{\bm{k}_1,\bm{k}_2,\bm{k}_3}\equiv{}&
\frac{i z^4}{\sqrt{2}}
\left(
\eta^{\mu\nu}(\bm{k}_1-\bm{k}_2)^\rho+\eta^{\nu\rho}(\bm{k}_2-\bm{k}_3)^\mu+\eta^{\rho\mu}(\bm{k}_3-\bm{k}_1)^\nu\right)
\\
V^{\mu\nu\rho\sigma}\equiv{}&{} \frac{i z^4}{2} \left(2\eta^{\mu\rho} \eta^{\nu\sigma}-\eta^{\mu\nu} \eta^{\rho\sigma}-\eta^{\mu\sigma} \eta^{\nu\rho}\right)\;,
\\
V^{\mu\nu\rho\sigma\lambda\kappa}_{\bm{k}_1, \bm{k}_2, \bm{k}_3}\equiv{}&{} \frac{z^8}{4}\left[
\left(
\bm{k}_2^\mu\bm{k}_3^\nu\eta^{\rho\lambda}\eta^{\sigma\kappa}-2\bm{k}_2^\mu\bm{k}_3^\rho\eta^{\nu\lambda}\eta^{\sigma\kappa}
\right)+\text{ permutations}
\right]
\ee
where the permutations in the graviton vertex are generated by the permutation group element $(\bm{k}_1\bm{k}_2\bm{k}_3)(ikm)(jln)$ in cycle notation.\footnote{See section 3.2.1. of \cite{Raju:2012zs} for the full contracted expression.}

At tree level, the expression for a gluon/graviton Witten diagram of $m$-external, $n$-propagators, $r$ 3-point vertices, and $s$ 4-point vertices reads as\footnote{At tree level, these quantities are not all independent and satisfy the equality $m+2n-3r-4s=0$.}$^{\text{,}}$\footnote{One can modify the graviton Witten diagram by adding higher point interactions as well, yet in this paper we stick to three point graviton interactions only.}
\bea[eq: generic tree level amplitude]
\label{eq: generic gluon tree level amplitude}
W^{\text{Tree}}_{\text{gluon}}=& \int_0^\infty
\frac{ dz_1}{z_1^{d+1}}\dots \frac{ dz_{r+s}}{z_{r+s}^{d+1}}
\prod\limits_{a=1}^{m} t_{\mu_a}(\tl z_a,k_a)
\prod\limits_{b=1}^{n}
\cG_{\nu_{2b-1}\nu_{2b}}(\widehat z_{2b-1},\widehat z_{2b},\bm{q}_{b})
\nn\\&\qquad
\x\prod\limits_{c=1,4,7,\dots}^{1+3(r-1)}
V^{\r_c\r_{c+1}\r_{c+2}}_{\bm{q}'_{c},\bm{q}'_{c+1},\bm{q}'_{c+2}}
\prod\limits_{d=1,5,9,\dots}^{1+4(s-1)}
V^{\r_{d}\r_{d+1}\r_{d+2}\r_{d+3}}
\\
\label{eq: generic graviton tree level amplitude}
W^{\text{Tree}}_{\text{graviton}}=& \int_0^\infty
\frac{ dz_1}{z_1^{d+1}}\dots \frac{ dz_{r}}{z_{r}^{d+1}}
\prod\limits_{a=1}^{m} t_{\mu_{2a-1}\mu_{2a}}(\tl z_a,k_a)
\prod\limits_{b=1}^{n}
\cG_{\nu_{4b-3}\nu_{4b-2}\nu_{4b-1}\nu_{4b}}(\widehat z_{2b-1},\widehat z_{2b},\bm{q}_{b})
\nn\\&\qquad
\x\prod\limits_{c=1,4,7,\dots}^{
	1+3(r-1)
}
V^{\r_{2c-1}\r_{2c}\r_{2c+1}\r_{2c+2}\r_{2c+3}\r_{2c+4}}_{\bm{q}'_{c},\bm{q}'_{c+1},\bm{q}'_{c+2}}
\eea 
where $\tl z_i,\widehat z_i\in \{z_1,\dots,z_{r+s}\}$  is determined by the topology of the diagram and where $\bm{q}_i$ and $\bm{q}'_i$ are linear combinations of vectors $\bm{k}_i$, again determined by the topology. Also, the sets $\{\mu\}\cup \{\nu\}$ and $\{\r\}$ are equivalent and they are contracted: the way which pairs are contracted depends on the topology of the diagram too.

For loop diagrams, the only new ingredient is the integration of  the loop momenta $\bm{l}$ at which the propagator momenta $\bm{q}$  and $\bm{q}'$ are now implicitly dependent; for a Witten diagram of $u-$loops, the expression simply reads as
\bea
\label{eq: gluon loop 1}
W^{\text{Loop}}_{\text{gluon}}= &
\prod\limits_{a=1}^{m}
\prod\limits_{b=1}^{n}
\prod\limits_{c=1,4,7,\dots}^{1+3(r-1)}
\prod\limits_{e=1,5,9,\dots}^{1+4(s-1)}
V^{\r_{e}\r_{e+1}\r_{e+2}\r_{e+3}}
\int_0^\infty
\frac{ dz_1}{z_1^{d+1}}\dots \frac{ dz_{r+s}}{z_{r+s}^{d+1}}
 t_{\mu_a}(\tl z_a,k_a)
\nn \\
 {}&\quad\x \prod\limits_{f=1}^u \int d^d\bm{l}_f
\cG_{\nu_{2b-1}\nu_{2b}}(\widehat z_{2b-1},\widehat z_{2b},\bm{q}_{b})
V^{\r_c\r_{c+1}\r_{c+2}}_{\bm{q}'_{c},\bm{q}'_{c+1},\bm{q}'_{c+2}}
\\
\label{eq: graviton loop 1}
W^{\text{Loop}}_{\text{graviton}}= &
\prod\limits_{a=1}^{m}
\prod\limits_{b=1}^{n}
\prod\limits_{c=1,4,7,\dots}^{1+3(r-1)}
\int_0^\infty
\frac{ dz_1}{z_1^{d+1}}\dots \frac{ dz_{r}}{z_{r}^{d+1}}
 t_{\mu_{2a-1}\mu_{2a}}(\tl z_a,k_a)
\nn\\ {}&\quad\x 
 \prod\limits_{f=1}^u \int d^d\bm{l}_f
\cG_{\mu_{4b-3}\mu_{4b-2}\mu_{4b-1}\mu_{4b}}(\widehat z_{2b-1},\widehat z_{2b},\bm{q}_{b})
V^{\nu_{2c-1}\nu_{2c}\nu_{2c+1}\nu_{2c+2}\nu_{2c+3}\nu_{2c+4}}_{\bm{q}'_{c},\bm{q}'_{c+1},\bm{q}'_{c+2}}
\eea

In \cite{Albayrak:2019asr, Albayrak:2019yve}, one insight to simplify the computation was to rewrite the propagators as differential operators acting on simpler propagators. Indeed, we observe that
\be 
\label{eq: propagators form 1}
\mathcal{G}_{\mu\nu}(\bm{k}; z, z')={}&{}\cD_{\mu\nu}^{\bm{k}}\int
p dp
\JJ{d-2}{d-2}{k,p}{z,z'}
\\
\mathcal{G}_{\mu\nu\rho\sigma}(\bm{k}; z, z')={}&{}\cD_{\mu\nu\rho\sigma}^{\bm{k}}\int
p dp
\JJ{d-4}{d}{k,p}{z,z'}
\ee 
for
\be 
\cD_{\mu\nu}^{\bm{k}}\equiv & \Pi_{\mu\nu}^{(1)\bm{k}}+\left( \Pi_{\mu\nu}^{(2)\bm{k}}\lim\limits_{k\rightarrow 0}\right)
\\
\cD_{\mu\nu\rho\sigma}^{\bm{k}}\equiv & \Pi_{\mu\nu\rho\sigma}^{(1)\bm{k}}+\left( \Pi_{\mu\nu\rho\sigma}^{(2)\bm{k}}\lim\limits_{k\rightarrow 0}\right)+\left( \Pi_{\mu\nu\rho\sigma}^{(3)\bm{k}}\lim\limits_{k\rightarrow 0}\partial_{k^2}\right)
\ee 
with which \equref{eq: generic tree level amplitude} become
\be
\label{eq: tree leven gluon and graviton Witten diagrams} 
W^{\text{Tree}}_{\text{gluon}}\sim &
\cD^{m,n,r}_\text{gluon}
\int_0^\infty
\frac{ dz_1}{z_1^{d+1}}\dots \frac{ dz_{r+s}}{z_{r+s}^{d+1}}
\left(\prod\limits_{c=1}^rz_c^4\right)
\left(
\prod\limits_{a=1}^{m}
\f^{d-2}_{d-2}(k_a,\tl z_a)
\right)
\left(
\prod\limits_{b=1}^{n}
\int
p_b dp_b
\JJ{d-2}{d-2}{q_b,p_b}{\widehat z_{2b-1},\widehat z_{2b}}
\right)
\\
W^{\text{Tree}}_{\text{graviton}}\sim &
\cD^{m,n,r}_\text{graviton}
\int_0^\infty
\frac{ dz_1}{z_1^{d+1}}\dots \frac{ dz_{r}}{z_{r}^{d+1}}
\left(\prod\limits_{c=1}^rz_c^8\right)
\left(
\prod\limits_{a=1}^{m}
\f^{d-4}_{d}(k_a,\tl z_a)
\right)
\left(
\prod\limits_{b=1}^{n}
\int
p_b dp_b
\JJ{d-4}{d}{q_b,p_b}{\widehat z_{2b-1},\widehat z_{2b}}
\right)
\ee
where we also used \equref{eq: bulk to boundary propagators in terms of scalar propagator}. Here the additional $z^{4,8}$ factors come from the $z-$dependence of three point vertices where the rest of the relevant factors are included in $\cD^{m,n,r}$.

The operator $\cD$ above consists of contraction of tensor structures in the Witten diagram but its details are not really important. The real importance of this form of the Witten diagram is that it drastically reduces the number of integrations because it generates the full Witten diagram by acting on a \emph{scalar factor} with a differential operator whose action simply consists of derivatives, limits, and contractions, all of which can be easily automated in a computer algebra program.  In contrast, symbolic integrations of interest here are computationally costly and reducing the total number of integrations enables the computations of higher order Witten diagrams in practice (see \cite{Albayrak:2018tam, Albayrak:2019asr, Albayrak:2019yve, Albayrak:2020isk} for further details with explicit results).

Once we move beyond tree level, the momenta $\bm{q}$ dependence of  $\cD^{\bm{q}}_{\mu\nu}$ and $\cD^{\bm{q}}_{\mu\nu\rho\sigma}$ spoils the nice separation of the scalar factor from the rest because we cannot take the differential operator outside the loop momenta $\bm{l}$ integral due to $\bm{l}$ dependence of  $\bm{q}$. To circumvent this problem, we present here an alternative representation for the propagators:
\be 
\label{eq: propagators form 2}
\mathcal{G}_{\mu\nu}(\bm{k}; z, z')={}&{}\tl\cD_{\mu\nu}^{(a)\bm{v}}\left[
\frac{(\bm{k}\.\bm{v})^2}{2!k^2}
\int
p dp
\left(
\frac{k^2+p^2}{p^2}
\right)^a
\JJ{d-2}{d-2}{k,p}{z,z'}
\right]
\\
\mathcal{G}_{\mu\nu\rho\sigma}(\bm{k}; z, z')={}&{}\tl\cD_{\mu\nu\rho\sigma}^{(a)\bm{v}}\left[
\frac{(\bm{k}\.\bm{v})^4}{4!k^4}\int
p dp
\left(
\frac{k^2+p^2}{p^2}
\right)^a
\JJ{d-4}{d}{k,p}{z,z'}
\right]
\ee 
for auxiliary polarization vectors $\bm{v}$, where we define
\be 
\label{eq: new differential operators}
\tl\cD^{(n)\bm{v}}_{\mu\nu}\equiv&\left(
\tl\Pi^{(1)\bm{v}}_{\mu\nu}\lim\limits_{n\rightarrow 0}
\right) +\left(\tl\Pi^{(2)\bm{v}}_{\mu\nu}\lim\limits_{n\rightarrow 1}\right)
\\
\tl\cD^{(n)\bm{v}}_{\mu\nu\rho\sigma}\equiv&\left(
\tl\Pi^{(1)\bm{v}}_{\mu\nu\rho\sigma}\lim\limits_{n\rightarrow 0}
\right) +\left(\tl\Pi^{(2)\bm{v}}_{\mu\nu\rho\sigma}\lim\limits_{n\rightarrow 1}\right)
+\left(\tl\Pi^{(3)\bm{v}}_{\mu\nu\rho\sigma}\lim\limits_{n\rightarrow 2}\right)
\ee 
in terms of  the modified projectors $\tl\Pi$. Likewise, we use these auxiliary vectors to rewrite the tensor structure of three point vertex factors to be independent of $\bm{k}$:
\be
\label{eq: modified vertex factors}
V^{\mu\nu\rho}_{\bm{k}_1,\bm{k}_2,\bm{k}_3}\equiv{}&
\tl V^{\mu\nu\rho}_{\bm{v}_1,\bm{v}_2,\bm{v}_3}\left[
\frac{i z^4}{\sqrt{2}}\sum\limits_{i=1}^3 (\bm{k}_i\.\bm{v}_i)
\right]
\\
V^{\mu\nu\rho\sigma\lambda\kappa}_{\bm{k}_1, \bm{k}_2, \bm{k}_3}\equiv{}&{} \tl V^{\mu\nu\rho\sigma\lambda\kappa}_{\bm{v}_1, \bm{v}_2, \bm{v}_3}\left[
\frac{z^8}{4}\sum\limits_{\substack{i=1,2,3\\j=i\text{ mod }3}} (\bm{k}_i\.\bm{v}_i)(\bm{k}_{j+1}\.\bm{v}_{j+1})
\right]
\ee
With these ingredients, we can rewrite \equref{eq: gluon loop 1} and \equref{eq: graviton loop 1} in a form similar to \equref{eq: tree leven gluon and graviton Witten diagrams}:
\be
\label{eq: diffrential form of loop level Witten diagrams}
W^{\text{Loop}}_{\text{gluon}}=\tl\cD^{m,n,r,s}_\text{gluon} \cM^{m,n,r,s}_{\text{gluon}}
\;,\quad 
W^{\text{Loop}}_{\text{graviton}}=\tl\cD^{m,n,r}_\text{graviton} \cM^{m,n,r}_{\text{graviton}}
\ee
where $\tl\cD$ carries all tensor structure information and where $\cM$ is simply a scalar factor. As $\tl\cD$ consists of derivatives, limits, and algebraic manipulations, it can be straightforwardly and efficiently applied once the scalar factor is known. On the other hand, scalar factor has all the integrations which are particularly challenging for symbolic arguments unless carried out at specific conditions (such as gluons in AdS$_4$). Therefore, in the rest of the paper, we will focus on scalar factors.

\section{Scalar factors for spinning Witten diagrams}
\label{sec: scalar factors}
The scalar factors  for loop level Witten diagrams defined in \equref{eq: diffrential form of loop level Witten diagrams} read as 
 \begin{subequations}
 	\label{eq: scalar factors}
\begin{multline}
\cM^{m,n,r,s}_{\text{gluon}}\equiv  \prod\limits_{f=1}^u \int d^d\bm{l}_f
\int_0^\infty
\frac{ dz_1}{z_1^{d+1}}\dots \frac{ dz_{r+s}}{z_{r+s}^{d+1}}
\left(
\prod\limits_{c=1,4,7,\dots}^{1+3(r-1)}
\frac{i \zb_c^4}{\sqrt{2}}\sum\limits_{i=c}^{c+2} (\bm{q}'_i\.\bm{v}'_i)
\right)
\left(
\prod\limits_{a=1}^{m}
\f^{d-2}_{d-2}(k_a,\tl z_a)
\right)
\\\x
\left(
\prod\limits_{b=1}^{n}
\frac{(\bm{q}_b\.\bm{v_b})^2}{2!q_b^2}
\int
p_b dp_b
\left(
\frac{q_b^2+p_b^2}{p_b^2}
\right)^{a_b}
\JJ{d-2}{d-2}{q_b,p_b}{\widehat z_{2b-1},\widehat z_{2b}}
\right)
\end{multline}
for gluons and
\begin{multline}
\cM^{m,n,r}_{\text{graviton}}\equiv \prod\limits_{f=1}^u \int d^d\bm{l}_f
\int_0^\infty
\frac{ dz_1}{z_1^{d+1}}\dots \frac{ dz_{r}}{z_{r}^{d+1}}
\left(
\prod\limits_{c=1,4,7,\dots}^{1+3(r-1)}
\frac{\zb_c^8}{4}\sum\limits_{\substack{i=0,1,2\\j=(i+1)\text{ mod }3}} (\bm{q}'_{c+i}\.\bm{v}'_{c+i})(\bm{q}'_{c+j}\.\bm{v}'_{c+j})
\right)
\\\x\left(
\prod\limits_{a=1}^{m}
\f^{d-4}_{d}(k_a,\tl z_a)
\right)
\left(
\prod\limits_{b=1}^{n}
\frac{(\bm{q}_b\.\bm{v}_b)^4}{4!q_b^4}\int
p_b dp_b
\left(
\frac{q_b^2+p_b^2}{p_b^2}
\right)^{a_b}
\JJ{d-4}{d}{q_b,p_b}{\widehat z_{2b-1},\widehat z_{2b}}
\right)
\end{multline}
\end{subequations}
for gravitons, where $\bm{q}_a$ (or $\bm{q}_a'$) is the momenta of the propagator $a$  whose dependence on the external momenta $\bm{k}_b$ and the loop momenta $\bm{l}_c$ is determined by the topology of the diagram at hand. Likewise, $\widehat z_a,\tl z_a,$ and $\zb_a$ are one of bulk points $z_i$, where topology determines which one they are.

\subsection{Examples: Bubble, triangle, and box gluon diagrams}

\begin{figure}
	\centering
\includegraphics[scale=1]{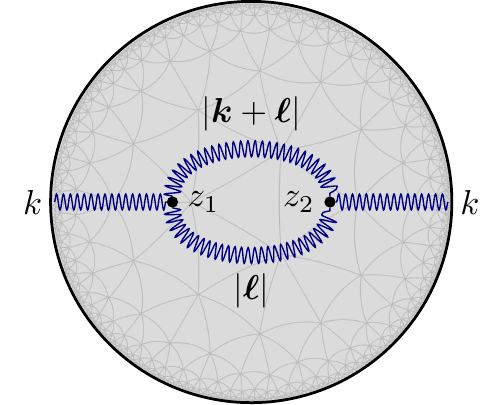}
\includegraphics[scale=1]{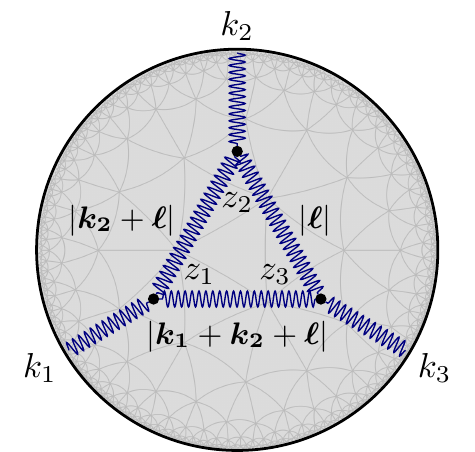}
\includegraphics[scale=1]{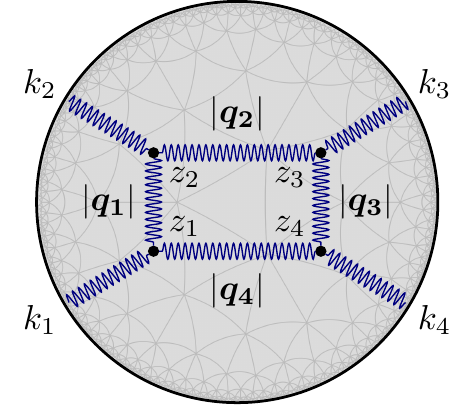}
\caption{\label{fig: gluon loop diagrams} Example of various gluon loop diagrams}
\end{figure}

Despite the complicated look of the general form in \equref{eq: scalar factors}, the scalar factors become simple  for particular Witten diagrams; for example, for the bubble diagram in \figref{\ref{fig: gluon loop diagrams}}, we have
\begin{multline}
\cM_{\substack{\text{gluon}\\\text{bubble}\\\text{diagram}}}=\cM^{2,2,2,0}_{\text{gluon}}\equiv  \int d^d\bm{l}
\int_0^\infty
\frac{ dz_1}{z_1^{d+1}}\frac{ dz_2}{z_2^{d+1}}
\frac{i z_1^4}{\sqrt{2}}
\frac{i z_2^4}{\sqrt{2}}
\left(
\sum\limits_{i=1}^{3} (\bm{q}'_i\.\bm{v}'_i)
\right)
\left(
\sum\limits_{i=4}^{6} (\bm{q}'_i\.\bm{v}'_i)
\right)
\\\x \f^{d-2}_{d-2}(k, z_1)
\f^{d-2}_{d-2}(k, z_2)
\left(
\frac{(\bm{l}\.\bm{v_1})^2}{2!l^2}
\int
p_1 dp_1
\left(
\frac{l^2+p_1^2}{p_1^2}
\right)^{a_1}
\JJ{d-2}{d-2}{l,p_1}{z_1,z_2}
\right)
\\\x
\left(
\frac{((\bm{k}+\bm{l})\.\bm{v_2})^2}{2!\abs{\bm{k}+\bm{l}}^2}
\int
p_2 dp_2
\left(
\frac{\abs{\bm{k}+\bm{l}}^2+p_2^2}{p_2^2}
\right)^{a_2}
\JJ{d-2}{d-2}{\abs{\bm{k}+\bm{l}},p_2}{z_1,z_2}
\right)
\end{multline}
which can be reorganized as
\begin{multline}
\label{eq: gluon bubble midstep}
\cM_{\substack{\text{gluon}\\\text{bubble}\\\text{diagram}}}=  -\frac{1}{8}
\int\limits_0^\infty
p_1dp_1p_2dp_2
\left(
\int_0^\infty z_1dz_1\f^{d-2}_{d-2}(k, z_1)J_\frac{d-2}{2}(p_1 z_1)J_\frac{d-2}{2}(p_2 z_1)
\right)
\\\x
\left(
 \int_0^\infty z_2 dz_2\f^{d-2}_{d-2}(k, z_2)J_\frac{d-2}{2} (p_1 z_2)J_\frac{d-2}{2} (p_2 z_2)
\right)
\\\x 
\left(\int d^d\bm{l}
\frac{(\bm{l}\.\bm{v_1})^2((\bm{k}+\bm{l})\.\bm{v_2})^2\left(
	\sum\limits_{i=1}^{3} (\bm{q}'_i\.\bm{v}'_i)
	\right)
	\left(
	\sum\limits_{i=4}^{6} (\bm{q}'_i\.\bm{v}'_i)
	\right)}{l^2\abs{\bm{k}+\bm{l}}^2\left(l^2+p_1^2-i\epsilon\right)\left(\abs{\bm{k}+\bm{l}}^2+p_2^2-i\epsilon\right)} \left(
\frac{l^2+p_1^2}{p_1^2}
\right)^{a_1}
\left(
\frac{\abs{\bm{k}+\bm{l}}^2+p_2^2}{p_2^2}
\right)^{a_2}
\right)
\end{multline}
where we can take
\be 
\bm{q}'_1=&\bm{q}'_4=\bm{k}\\
\bm{q}'_2=&\bm{q}'_5=\bm{l}\\
\bm{q}'_3=&\bm{q}'_6=\bm{k}+\bm{l}
\ee
Similarly, we can write down the scalar factors associated with the triangle and box diagrams as follows.
\begin{subequations}
\label{eq: gluon triangle and box diagrams}
\begin{multline}
\cM_{\substack{\text{gluon}\\\text{triangle}\\\text{diagram}}}=\cM^{3,3,3,0}_{\text{gluon}}=  \left(\frac{i}{\sqrt{2}}\right)^3\int d^d\bm{l}_f
\int_0^\infty
\frac{dz_1dz_2dz_3}{(z_1z_2z_3)^{d-3}}
\left(
\sum\limits_{i=1}^{3} (\bm{q}'_i\.\bm{v}'_i)
\right)
\\\x
\left(
\sum\limits_{i=4}^{6} (\bm{q}'_i\.\bm{v}'_i)
\right)
\left(
\sum\limits_{i=7}^{9} (\bm{q}'_i\.\bm{v}'_i)
\right)\left(\prod_{i=1}^3\f^{d-2}_{d-2}(k_i, z_i)\right)
\\\x
\left(
\frac{\left(\left(\bm{k_2}+\bm{l}\right)\.\bm{v_1}\right)^2}{2!\abs{\bm{k_2}+\bm{l}}^2}
\int
p_1 dp_1
\left(
\frac{\abs{\bm{k_2}+\bm{l}}^2+p_1^2}{p_1^2}
\right)^{a_1}
\JJ{d-2}{d-2}{\abs{\bm{k_2}+\bm{l}},p_1}{z_1,z_2}
\right)
\\\x
\left(
\frac{(\bm{l}\.\bm{v_2})^2}{2!l^2}
\int
p_2 dp_2
\left(
\frac{l^2+p_2^2}{p_2^2}
\right)^{a_2}
\JJ{d-2}{d-2}{l,p_2}{z_2,z_3}
\right)
\\\x
\left(
\frac{\left(\left(\bm{k_1}+\bm{k_2}+\bm{l}\right)\.\bm{v_3}\right)^2}{2!\abs{\bm{k_1}+\bm{k_2}+\bm{l}}^2}
\int
p_3 dp_3
\left(
\frac{\abs{\bm{k_1}+\bm{k_2}+\bm{l}}^2+p_3^2}{p_3^2}
\right)^{a_3}
\JJ{d-2}{d-2}{\abs{\bm{k_1}+\bm{k_2}+\bm{l}},p_3}{z_3,z_1}
\right)
\end{multline}
for
\be 
\bm{q}'_1=&\bm{k}_1 &\qquad \bm{q}'_5=\bm{q}'_8=&\bm{l}\\
\bm{q}'_4=&\bm{k}_2&\bm{q}'_2=\bm{q}'_6=&\bm{k}_2+\bm{l}\\
\bm{q}'_7=&\bm{k}_3&\bm{q}'_3=\bm{q}'_9=&\bm{k}_1+\bm{k}_2+\bm{l}
\ee 
\end{subequations}
and
\begin{subequations}
\begin{multline}
\cM_{\substack{\text{gluon}\\\text{box}\\\text{diagram}}}=\cM^{4,4,4,0}_{\text{gluon}}=  \left(\frac{i}{\sqrt{2}}\right)^4\int d^d\bm{l}_f
\int_0^\infty
\frac{dz_1dz_2dz_3dz_4}{(z_1z_2z_3z_4)^{d-3}}
\left(
\sum\limits_{i=1}^{3} (\bm{q}'_i\.\bm{v}'_i)
\right)
\\\x
\left(
\sum\limits_{i=4}^{6} (\bm{q}'_i\.\bm{v}'_i)
\right)
\left(
\sum\limits_{i=7}^{9} (\bm{q}'_i\.\bm{v}'_i)
\right)
\left(
\sum\limits_{i=10}^{12} (\bm{q}'_i\.\bm{v}'_i)
\right)
\left(\prod_{i=1}^4\f^{d-2}_{d-2}(k_i, z_i)\right)
\\\x
\left(
\frac{\left(\left(\bm{k_2}+\bm{l}\right)\.\bm{v_1}\right)^2}{2!\abs{\bm{k_2}+\bm{l}}^2}
\int
p_1 dp_1
\left(
\frac{\abs{\bm{k_2}+\bm{l}}^2+p_1^2}{p_1^2}
\right)^{a_1}
\JJ{d-2}{d-2}{\abs{\bm{k_2}+\bm{l}},p_1}{z_1,z_2}
\right)
\\\x
\left(
\frac{(\bm{l}\.\bm{v_2})^2}{2!l^2}
\int
p_2 dp_2
\left(
\frac{l^2+p_2^2}{p_2^2}
\right)^{a_2}
\JJ{d-2}{d-2}{l,p_2}{z_2,z_3}
\right)
\\\x
\left(
\frac{\left(\left(\bm{l}-\bm{k_3}\right)\.\bm{v_3}\right)^2}{2!\abs{\bm{l}-\bm{k_3}}^2}
\int
p_3 dp_3
\left(
\frac{\abs{\bm{l}-\bm{k_3}}^2+p_3^2}{p_3^2}
\right)^{a_3}
\JJ{d-2}{d-2}{\abs{\bm{l}-\bm{k_3}},p_3}{z_3,z_4}
\right)
\\\x
\left(
\frac{\left(\left(\bm{k_1}+\bm{k_2}+\bm{l}\right)\.\bm{v_4}\right)^2}{2!\abs{\bm{k_1}+\bm{k_2}+\bm{l}}^2}
\int
p_4 dp_4
\left(
\frac{\abs{\bm{k_1}+\bm{k_2}+\bm{l}}^2+p_4^2}{p_4^2}
\right)^{a_4}
\JJ{d-2}{d-2}{\abs{\bm{k_1}+\bm{k_2}+\bm{l}},p_4}{z_4,z_1}
\right)
\end{multline}
for
\be 
\bm{q}'_1=&\bm{k}_1 &\qquad \bm{q}'_5=\bm{q}'_8=&\bm{l}\\
\bm{q}'_4=&\bm{k}_2&\bm{q}'_2=\bm{q}'_6=&\bm{k}_2+\bm{l}\\
\bm{q}'_7=&\bm{k}_3&\bm{q}'_9=\bm{q}'_{11}=&\bm{l}-\bm{k}_3\\
\bm{q}'_{10}=&\bm{k}_4&\bm{q}'_3=\bm{q}'_{12}=&\bm{k}_1+\bm{k}_2+\bm{l}
\ee 
\end{subequations}
\subsection{Computing bubble diagram}
Let us recall the scalar factor for bubble diagram from \equref{eq: gluon bubble midstep}:
\begin{multline}
\label{eq: gluon bubble midstep 3}
\cM_{\substack{\text{gluon}\\\text{bubble}\\\text{diagram}}}=  -\frac{1}{8}
\int\limits_0^\infty
p_1dp_1p_2dp_2
\left(
\int_0^\infty zdz\f^{d-2}_{d-2}(k, z)J_\frac{d-2}{2}(p_1 z)J_\frac{d-2}{2}(p_2 z)
\right)^2
\\\x 
\left(\int d^d\bm{l}
\frac{(\bm{l}\.\bm{v_1})^2((\bm{k}+\bm{l})\.\bm{v_2})^2\left(
	\sum\limits_{i=1}^{3} (\bm{q}'_i\.\bm{v}'_i)
	\right)
	\left(
	\sum\limits_{i=4}^{6} (\bm{q}'_i\.\bm{v}'_i)
	\right)}{l^2\abs{\bm{k}+\bm{l}}^2\left(l^2+p_1^2-i\epsilon\right)\left(\abs{\bm{k}+\bm{l}}^2+p_2^2-i\epsilon\right)}\left(
\frac{l^2+p_1^2}{p_1^2}
\right)^{a_1}
\left(
\frac{\abs{\bm{k}+\bm{l}}^2+p_2^2}{p_2^2}
\right)^{a_2}
\right)
\end{multline}
for
\be 
\bm{q}'_1=&\bm{q}'_4=\bm{k}\\
\bm{q}'_2=&\bm{q}'_5=\bm{l}\\
\bm{q}'_3=&\bm{q}'_6=\bm{k}+\bm{l}
\ee
The first piece in \equref{eq: gluon bubble midstep 3} can be computed analytically in terms of  Appell's hypergeometric functions:\footnote{Please see \secref{\ref{sec: generic three point interactions}} for further details.}
\be 
\int_0^\infty zdz\f^{d-2}_{d-2}(k, z)J_\frac{d-2}{2}(p_1 z)J_\frac{d-2}{2}(p_2 z)
=
\frac{2^{\frac{3d-5}{2}}\Gamma\left(\frac{d-1}{2}\right)\left(\frac{p_1p_2}{k^2}\right)^{\frac{d-2}{2}}}{\pi k^2} F_4\left(\frac{d}{2},d-1;\frac{d}{2},\frac{d}{2};-\frac{p_1^2}{k^2},-\frac{p_2^2}{k^2}\right)
\ee 
hence we have
\begin{multline}
\label{eq: gluon bubble midstep 2}
\cM_{\substack{\text{gluon}\\\text{bubble}\\\text{diagram}}}=  -
\frac{2^{3d-8}\Gamma\left(\frac{d-1}{2}\right)^2}{\pi^2 k^{2d}}
\int\limits_0^\infty
dp_1dp_2  p_1^{d-1-2a_1}p_2^{d-1-2a_2}
\left[
F_4\left(\frac{d}{2},d-1;\frac{d}{2},\frac{d}{2};-\frac{p_1^2}{k^2},-\frac{p_2^2}{k^2}\right)
\right]^2
\\\x 
\left(\int d^d\bm{l}
\frac{(\bm{l}\.\bm{v_1})^2((\bm{k}+\bm{l})\.\bm{v_2})^2\left(
	\sum\limits_{i=1}^{3} (\bm{q}'_i\.\bm{v}'_i)
	\right)
	\left(
	\sum\limits_{i=4}^{6} (\bm{q}'_i\.\bm{v}'_i)
	\right)}{l^2\abs{\bm{k}+\bm{l}}^2\left(l^2+p_1^2-i\epsilon\right)^{1-a_1}\left(\abs{\bm{k}+\bm{l}}^2+p_2^2-i\epsilon\right)^{1-a_2}}
\right)
\end{multline}
which we can rewrite using the definition of $\bm{q}'$ above as
\begin{multline}
\cM_{\substack{\text{gluon}\\\text{bubble}\\\text{diagram}}}=  -
\frac{2^{3d-8}\Gamma\left(\frac{d-1}{2}\right)^2}{\pi^2 k^{2d}}
\int\limits_0^\infty
dp_1dp_2  p_1^{d-1-2a_1}p_2^{d-1-2a_2}
\left[
F_4\left(\frac{d}{2},d-1;\frac{d}{2},\frac{d}{2};-\frac{p_1^2}{k^2},-\frac{p_2^2}{k^2}\right)
\right]^2
\\\x 
\Bigg(
\cJ^{v_1',v_4',v_3',v_6',v_2,v_2}_{v_1,v_1,v_2',v_5'}
+\Bigg\{
2\cJ^{v_1',v_4',v_3',v_6',v_2}_{v_1,v_1,v_2,v_2',v_5'}
+\cJ^{v_1',v_4',v_3',v_2,v_2}_{v_1,v_1,v_2',v_5',v_6'}
+\cJ^{v_1',v_4',v_2,v_2,v_6'}_{v_1,v_1,v_2',v_5',v_3'}
\Bigg\}
\\
+\Bigg\{
\cJ^{v_1',v_4',v_3',v_6'}_{v_1,v_1,v_2,v_2,v_2',v_5'}
+2\cJ^{v_1',v_4',v_2,v_3'}_{v_1,v_1,v_2,v_2',v_5',v_6'}
+2\cJ^{v_1',v_4',v_2,v_6'}_{v_1,v_1,v_2,v_2',v_5',v_3'}
+\cJ^{v_1',v_4',v_2,v_2}_{v_1,v_1,v_2',v_5',v_3',v_6'}
\Bigg\}
\\
+\Bigg\{\cJ^{v_1',v_4',v_3'}_{v_1,v_1,v_2,v_2,v_2',v_5',v_6'}
+\cJ^{v_1',v_4',v_6'}_{v_1,v_1,v_2,v_2,v_2',v_5',v_3'}
+2\cJ^{v_1',v_4',v_2}_{v_1,v_1,v_2,v_2',v_5',v_3',v_6'}
\Bigg\}
+\cJ^{v_1',v_4'}_{v_1,v_1,v_2,v_2,v_2',v_5',v_3',v_6'}
\Bigg)
\end{multline}
where we have defined

\begin{multline}
\mathcal{J}^{a_1,a_2,\dots,a_m}_{b_1,b_2,\dots,b_n}\equiv (\bm{k}\.\bm{a_1})(\bm{k}\.\bm{a_2})\cdots 
\left(\bm{k}\.\bm{a_m}\right)\\\x
\int d^d\bm{l}
\frac{(\bm{l}\.\bm{b_1})(\bm{l}\.\bm{b_2})\cdots 
	\left(\bm{l}\.\bm{b_n}\right)
}{l^2\abs{\bm{k}+\bm{l}}^2\left(l^2+p_1^2-i\epsilon\right)^{1-a_1}\left(\abs{\bm{k}+\bm{l}}^2+p_2^2-i\epsilon\right)^{1-a_2}}
\end{multline}
for convenience.

Evaluation of  $\mathcal{J}^{a_1,a_2,\dots,a_m}_{b_1,b_2,\dots,b_n}$ for generic $a_{1,2}$ is somewhat complicated, however we can simplify it by noting that only $a_1,a_2=0,1$ are relevant which can be checked through \equref{eq: diffrential form of loop level Witten diagrams}, \equref{eq: new differential operators} and \equref{eq: definition of d tilde}. Therefore, we can make the replacement
\begin{multline}
\mathcal{J}^{a_1,a_2,\dots,a_m}_{b_1,b_2,\dots,b_n}\rightarrow (\bm{k}\.\bm{a_1})(\bm{k}\.\bm{a_2})\cdots 
\left(\bm{k}\.\bm{a_m}\right)
\\\x\Bigg(\delta_{a_1}^0\delta_{a_2}^0
\int d^d\bm{l}
\frac{(\bm{l}\.\bm{b_1})(\bm{l}\.\bm{b_2})\cdots 
	\left(\bm{l}\.\bm{b_n}\right)
}{l^2\abs{\bm{k}+\bm{l}}^2\left(l^2+p_1^2-i\epsilon\right)\left(\abs{\bm{k}+\bm{l}}^2+p_2^2-i\epsilon\right)}
\\+\delta_{a_1}^0\delta_{a_2}^1
\int d^d\bm{l}
\frac{(\bm{l}\.\bm{b_1})(\bm{l}\.\bm{b_2})\cdots 
	\left(\bm{l}\.\bm{b_n}\right)
}{l^2\abs{\bm{k}+\bm{l}}^2\left(l^2+p_1^2-i\epsilon\right)}
\\+\delta_{a_1}^1\delta_{a_2}^0
\int d^d\bm{l}
\frac{(\bm{l}\.\bm{b_1})(\bm{l}\.\bm{b_2})\cdots 
	\left(\bm{l}\.\bm{b_n}\right)
}{l^2\abs{\bm{k}+\bm{l}}^2\left(\abs{\bm{k}+\bm{l}}^2+p_2^2-i\epsilon\right)}
\\+\delta_{a_1}^1\delta_{a_2}^1
\int d^d\bm{l}
\frac{(\bm{l}\.\bm{b_1})(\bm{l}\.\bm{b_2})\cdots 
	\left(\bm{l}\.\bm{b_n}\right)
}{l^2\abs{\bm{k}+\bm{l}}^2}
\Bigg)
\end{multline}
therefore the scalar factor for the gluon bubble diagram becomes sum of 48 terms, i.e.
\begin{multline}
\label{eq: gluon bubble diagram middle step}
\cM_{\substack{\text{gluon}\\\text{bubble}\\\text{diagram}}}= -\delta_{a_1}^1\delta_{a_2}^1
\left(\bm{k}\.\bm{v}_1'\right)
\left(\bm{k}\.\bm{v}_2\right)^2
\left(\bm{k}\.\bm{v}_3'\right)
\left(\bm{k}\.\bm{v}_4'\right)
\left(\bm{k}\.\bm{v}_6'\right)
\frac{2^{3d-8}\Gamma\left(\frac{d-1}{2}\right)^2}{\pi^2 k^{2d}}
\\\x 
\int\limits_0^\infty
\frac{dp_1dp_2}{(p_1p_2)^{3-d}}  
\left[
F_4\left(\frac{d}{2},d-1;\frac{d}{2},\frac{d}{2};-\frac{p_1^2}{k^2},-\frac{p_2^2}{k^2}\right)
\right]^2
\int d^d\bm{l}
\frac{(\bm{l}\.\bm{v}_1)^2(\bm{l}\.\bm{v}_2')\left(\bm{l}\.\bm{v}_5'\right)
}{l^2\abs{\bm{k}+\bm{l}}^2}
\\ +\text{ other terms}
\end{multline}

In Appendix~\ref{sec: loop integrals} we go over how to do such volume integrals in great generality via standard QFT tricks; the final result in \equref{eq: generic volume integral} reduces such involved integrals into various products, summations, $1d$ definite integrals of rational functions, and set-partitioning, all of which can be efficiently implemented in an algorithmic way in any computer computation software such as \texttt{Mathematica}. Indeed, we can rewrite \equref{eq: gluon bubble diagram middle step} with \equref{eq: volume integration example} as
\small 
\begin{multline}
\label{eq: gluon bubble diagram middle step 2}
\cM_{\substack{\text{gluon}\\\text{bubble}\\\text{diagram}}}= -\delta_{a_1}^1\delta_{a_2}^1\mathfrak{t}_{v_i,v_i'}^k
\frac{i^{d+1} (d+4) \Gamma \left(-\frac{d}{2}\right)
	\Gamma (d+3)}{k^{d-6}2^{8-d}\pi^{1-\frac{d}{2}}\left(d^2-1\right)^2}
\int\limits_0^\infty
\frac{dp_1dp_2}{(p_1p_2)^{3-d}}  
\left[
F_4\left(\frac{d}{2},d-1;\frac{d}{2},\frac{d}{2};-\frac{p_1^2}{k^2},-\frac{p_2^2}{k^2}\right)
\right]^2
\\ +\text{ other terms}
\end{multline}
\normalsize
where $\mathfrak{t}_{v_i,v_i'}^k$ is the overall tensor structure.\footnote{Its explicit form reads as
\begin{multline}
\mathfrak{t}_{v_i,v_i'}^k=
\frac{\left(\bm{k}\.\bm{v}_1'\right)
	\left(\bm{k}\.\bm{v}_2\right)^2
	\left(\bm{k}\.\bm{v}_3'\right)
	\left(\bm{k}\.\bm{v}_4'\right)
	\left(\bm{k}\.\bm{v}_6'\right)}{k^6}
\Bigg(
k^{-4}\left(\bm{k}\.\bm{v}_1\right)^{2}\left(\bm{k}\.\bm{v}_2'\right)\left(\bm{k}\.\bm{v}_5'\right)
\\
-\frac{d k^{-2}}{d+4}\Big[\left(\bm{k}\.\bm{v}_1\right)^2\left(\bm{v}_2'\.\bm{v}_5'\right)
+
2\left(\bm{k}\.\bm{v}_1\right)\left(\bm{k}\.\bm{v}_2'\right)\left(\bm{v}_1\.\bm{v}_5'\right)
+2
\left(\bm{k}\.\bm{v}_1\right)\left(\bm{k}\.\bm{v}_5'\right)\left(\bm{v}_1\.\bm{v}_2'\right)
\\+
\left(\bm{k}\.\bm{v}_2'\right)\left(\bm{k}\.\bm{v}_5'\right)\left(\bm{v}_1\.\bm{v}_1\right)
\Big]
+\frac{d}{d+4}\Big[\left(\bm{v}_1\.\bm{v}_1\right)\left(\bm{v}_2'\.\bm{v}_5'\right)+2\left(\bm{v}_1\.\bm{v}_2'\right)\left(\bm{v}_1\.\bm{v}_5'\right)\Big]
\Bigg)
\end{multline}
}

The other terms in the equation above are of similar form as well: they will simply have different overall-tensor-structure, and they may bring additional $p$ dependent terms inside the integration; however all of them can be computed using the same equation, that is \equref{eq: generic volume integral}. 

The remaining computation in \equref{eq: gluon bubble diagram middle step 2} is intricate which involves integrating products of hypergeometric functions, hence it is not sagacious to
insist to work in non-specific dimensions. However, the expression is very simple for specific $d$ values; for example,
\be 
F_4\left(\frac{d}{2},d-1;\frac{d}{2},\frac{d}{2};-\frac{p_1^2}{k^2},-\frac{p_2^2}{k^2}\right)\evaluated_{\substack{d=2n+1\\n\in\mathbb{N}}}=\left(\frac{k^4}{k^4+2k^2(p_1^2+p_2^2)+(p_1^2-p_2^2)^2}\right)^n
\ee 
with which the integration becomes doable with an appropriate regularization at any given $n$.

In summary, we observe that the loop-level computations become tractable in momentum space in AdS$_{d+1}$. Although we only illustrated the case for the gluons, the situation is similar for gravitons as well; what is common in both cases though is the very technical nature of the formalism that we unpacked above. However, the key point is that the computations in each and every step is algorithmic  and can be efficiently implemented in a computer computation software. In particular, momentum space formalism along with the way we decompose the Witten diagrams into differential operators and scalar factors effectively converts a mathematically hard problem into technical yet computer-friendly computation as the final result is simply derivatives and limits acting on a scalar factor which itself is computed via products, sums, and list partitioning, and all of these can be efficiently computed unlike a convoluted volume integral! The main result of the paper is  therefore the following prescription:
\begin{enumerate}
	\item For any given Witten diagram, rewrite it as $W=\tl\cD \cM$ where the differential operator $\tl\cD$ is given in \equref{eq: definition of d tilde} and $\cM$ in \equref{eq: scalar factors}.
	\item Unpack $\cM$ depending on the topology of chosen Witten diagram as is done in \equref{eq: gluon triangle and box diagrams} for gluon triangle and box diagrams.
	\item Rewrite the scalar factor such that it becomes of the form 
	\begin{equation*}
	\cM=
	\int dp_1dp_2\dots dp_m \left(\int dz_1\cdots\right)\cdots \left(\int dz_n\cdots\right)\left(\int d^dl_1\cdots\right)\cdots \left(\int d^dl_r\cdots\right)
	\end{equation*}
	which can always be done in the current formalism (see \equref{eq: gluon bubble midstep} as an example of this in case of gluon bubble diagram).
	\item Replace radial integrals of the AdS ($z$-integrations) in terms of Appell's $F_4$ functions, as is detailed in \secref{\ref{sec: generic three point interactions}}.
	\item Replace $l-$integrations in $\cM$ as given in. \equref{eq: generic volume integral}.\footnote{The $u-$integrations in \equref{eq: generic volume integral} can be immediately carried out for numeric $d$ values, but are not generically doable if we keep dimension symbolic.}
	\item With the replacements in the steps above, $\cM$ becomes summation of bunch of terms which involve products, summations, list partitioning, and $p-$integrations. In odd $d$ (such as the case for AdS$_4$), the Appell's $F_4$ function becomes meromorphic  in $p$ hence the $p-$integrations become straightforward (upto possible regularization).\footnote{It is an open question how one should proceed for even $d$. We believe it may be more efficient to compute the Witten diagrams case by case for even $d$, contrary to our generic approach in this paper. Of course, our formalism is perfectly fine and would be extremely generic if one could compute (or bypass) $p-$integrations of Appell's $F_4$ functions.}
	\item Apply the differential operator $\tl\cD$ to the scalar factor $\cM$ to obtain the full Witten diagram: as this merely amounts to taking derivatives and limits of a factor composed of summations, products, and list partitioning; all of these steps can be efficiently done algorithmically.
\end{enumerate}

\section{Conclusion}
In this paper, we have studied a formalism to compute loop amplitudes in Anti-de Sitter space  in Fourier space for gauge theory and gravity loops in AdS$_{d+1}$. In particular, we have constructed a differential operator which can act on a scalar factor to yield both Yang Mills and gravity loop correlators. In addition, we have presented a prescription which can be automated in order to perform tensorial loop computations in Anti-de Sitter space. There are myriad of interesting directions that one can pursue and we will list a few.

One of the main motivation of our work is to take the first step to connect AdS loops with cascading number of new ideas and techniques that are emerging in flat space.  For instance, in \cite{Bourjaily:2019exo}, it was shown that $n$- particle massive Feynman integrals in arbitrary dimensions of spacetime have nice geometric properties such as the connections with hyperbolic simplicial geometry and the answer respects dual conformal symmetry. This method can be directly applied to the computation of the above-mentioned AdS scale factor. Furthermore, we want to stress that we are motivated to study gluons and gravitons in AdS as many of the extremely powerful physical insights and mathematical structures in the last decade have occurred in the study of the flat space S-matrix of gauge theory and gravity \cite{Arkani-Hamed:2013jha}. It is tempting to contemplate if there are analogous geometric structures like the amplituhedron that exist for loop amplitudes in Anti-de Sitter space.

Similarly, as in the context of Minkowski space, AdS loops can also be expressed in terms of the special classes of multiple polylogarithms. In the context of flat space, there has been progress in demonstrating that these complicated polylogs can admit a much simpler analytic expression.  The technology used is called the \emph{symbol} map and this map can capture combinatorial and analytical properties of the complicated Feynman integrals \cite{Duhr:2011zq}. In a related work \cite{Hillman:2019wgh}, \emph{symbols} were used to compute loop amplitudes in de Sitter space. It would be natural to use these methods in the context of AdS loops. Likewise, it would be intriguing to incorporate cutting rules in momentum space AdS in the study of gluons and gravitons, and we are hoping to address it in a future work.

\section*{Acknowledgement}
SA and SK thank Chandramouli Chowdhury and David Meltzer for helpful discussions. The research of SA is supported by DOE grant no. DE-SC0020318 and Simons Foundation grant 488651. SK was supported by DRFC Discretionary Funds from Williams College.

\appendix
\section{Technical details}
\subsection{Projectors and differential operators}
\label{sec: projectors and differential operators}

In this appendix, we collect some of the technical details we skipped in main body. We first note the definition of the projectors $\Pi$ used in \equref{eq: traditional propagators}:
\be 
\Pi_{\mu\nu\rho\sigma}^{(1)\bm{k}} = &\frac{i}{2}\left(
\Pi^{(1)\bm{k}}_{\mu\rho}\Pi^{(1)\bm{k}}_{\nu\sigma}+\Pi^{(1)\bm{k}}_{\mu\sigma}\Pi^{(1)\bm{k}}_{\nu\rho}-\frac{2}{d-1}\Pi^{(1)\bm{k}}_{\mu\nu}\Pi^{(1)\bm{k}}_{\rho\sigma}\right)
\\
\Pi_{\mu\nu\rho\sigma}^{(2)\bm{k}} = &\frac{i}{2}\left(
\Pi^{(1)\bm{k}}_{\mu\rho}\Pi^{(2)\bm{k}}_{\nu\sigma}+\Pi^{(1)\bm{k}}_{\mu\sigma}\Pi^{(2)\bm{k}}_{\nu\rho}-\frac{2}{d-1}\Pi^{(1)\bm{k}}_{\mu\nu}\Pi^{(2)\bm{k}}_{\rho\sigma}
\right)
\\&+\frac{i}{2}\left(
\Pi^{(2)\bm{k}}_{\mu\rho}\Pi^{(1)\bm{k}}_{\nu\sigma}+\Pi^{(2)\bm{k}}_{\mu\sigma}\Pi^{(1)\bm{k}}_{\nu\rho}-\frac{2}{d-1}\Pi^{(2)\bm{k}}_{\mu\nu}\Pi^{(1)\bm{k}}_{\rho\sigma}\right)
\\&+\frac{i}{2}\left(
\Pi^{(2)\bm{k}}_{\mu\rho}\Pi^{(2)\bm{k}}_{\nu\sigma}+\Pi^{(2)\bm{k}}_{\mu\sigma}\Pi^{(2)\bm{k}}_{\nu\rho}-\frac{2}{d-1}\Pi^{(2)\bm{k}}_{\mu\nu}\Pi^{(2)\bm{k}}_{\rho\sigma}
\right)
\\
\Pi_{\mu\nu\rho\sigma}^{(3)\bm{k}} = &\frac{i}{2}\left(
\Pi^{(2)\bm{k}}_{\mu\rho}\Pi^{(2)\bm{k}}_{\nu\sigma}+\Pi^{(2)\bm{k}}_{\mu\sigma}\Pi^{(2)\bm{k}}_{\nu\rho}-\frac{2}{d-1}\Pi^{(2)\bm{k}}_{\mu\nu}\Pi^{(2)\bm{k}}_{\rho\sigma}\right)
\ee 
and
\begin{equation}
\label{eq: projectors}
\Pi^{(1)\bm{k}}_{\mu\nu}\equiv{}\frac{\eta_{\mu\nu}k^2-\bm{k}_\mu\bm{k}_\nu}{ik^2}\;,\quad 
\Pi^{(2)\bm{k}}_{\mu\nu}\equiv{}{}\frac{\bm{k}_\mu\bm{k}_\nu}{ik^2}\;.
\end{equation}
We likewise note the definition of  the differential operators in \equref{eq: tree leven gluon and graviton Witten diagrams}:
\be 
\cD^{m,n,r}_\text{gluon}\equiv &
\left(
\prod\limits_{c=1,4,7,\dots}^{1+3(r-1)}
\dot V^{\r_c\r_{c+1}\r_{c+2}}_{\bm{q}'_{c},\bm{q}'_{c+1},\bm{q}'_{c+2}}
\right)
\left(
\prod\limits_{e=1,5,9,\dots}^{m+2n-3r-3}
V^{\r_{e}\r_{e+1}\r_{e+2}\r_{e+3}}
\right)
\left(
\prod\limits_{a=1}^{m}
\bm{\epsilon}^a_{\mu_a}
\right)
\left(
\prod\limits_{b=1}^{n}
\cD_{\nu_{2b-1}\nu_{2b}}^{\bm{q}_b}
\right)
\\
\cD^{m,n,r}_\text{graviton}\equiv &
\left(
\prod\limits_{c=1,4,7,\dots}^{m+2n-2}\dot V^{\r_{2c-1}\r_{2c}\r_{2c+1}\r_{2c+2}\r_{2c+3}\r_{2c+4}}_{\bm{q}'_{c},\bm{q}'_{c+1},\bm{q}'_{c+2}}
\right)
\left(
\prod\limits_{a=1}^{m}
\bm{\epsilon}^a_{\mu_{2a-1}\mu_{2a}}
\right)
\left(
\prod\limits_{b=1}^{n}
\cD_{\nu_{4b-3}\nu_{4b-2}\nu_{4b-1}\nu_{4b}}^{\bm{q}_b}
\right)
\ee 
where three point vertex factors $\dot V$ are $V$ with their $z-$dependencies stripped off!

The modified projectors for gluons are given as follows
\be 
\tl\Pi^{(1)\bm{v}}_{\mu\nu}\equiv -i\left(\eta_{\mu\nu}\pdr{}{\bm{v}^\rho}\pdr{}{\bm{v}_\rho}-\pdr{}{\bm{v}^\mu}\pdr{}{\bm{v}^\nu}\right)\;,\quad \tl\Pi^{(2)\bm{v}}_{\mu\nu}\equiv -i\pdr{}{\bm{v}^\mu}\pdr{}{\bm{v}^\nu}
\ee 
and the modified projectors for gravitons are defined in terms of them:
\be 
\tl\Pi_{\mu\nu\rho\sigma}^{(1)\bm{k}} = &\frac{i}{2}\left(
\tl\Pi^{(1)\bm{k}}_{\mu\rho}\tl\Pi^{(1)\bm{k}}_{\nu\sigma}+\tl\Pi^{(1)\bm{k}}_{\mu\sigma}\tl\Pi^{(1)\bm{k}}_{\nu\rho}-\frac{2}{d-1}\tl\Pi^{(1)\bm{k}}_{\mu\nu}\tl\Pi^{(1)\bm{k}}_{\rho\sigma}\right)
\\
\tl\Pi_{\mu\nu\rho\sigma}^{(2)\bm{k}} = &\frac{i}{2}\left(
\tl\Pi^{(1)\bm{k}}_{\mu\rho}\tl\Pi^{(2)\bm{k}}_{\nu\sigma}+\tl\Pi^{(1)\bm{k}}_{\mu\sigma}\tl\Pi^{(2)\bm{k}}_{\nu\rho}-\frac{2}{d-1}\tl\Pi^{(1)\bm{k}}_{\mu\nu}\tl\Pi^{(2)\bm{k}}_{\rho\sigma}
\right)
\\&+\frac{i}{2}\left(
\tl\Pi^{(2)\bm{k}}_{\mu\rho}\tl\Pi^{(1)\bm{k}}_{\nu\sigma}+\tl\Pi^{(2)\bm{k}}_{\mu\sigma}\tl\Pi^{(1)\bm{k}}_{\nu\rho}-\frac{2}{d-1}\tl\Pi^{(2)\bm{k}}_{\mu\nu}\tl\Pi^{(1)\bm{k}}_{\rho\sigma}\right)
\\
\tl\Pi_{\mu\nu\rho\sigma}^{(3)\bm{k}} = &\frac{i}{2}\left(
\tl\Pi^{(2)\bm{k}}_{\mu\rho}\tl\Pi^{(2)\bm{k}}_{\nu\sigma}+\tl\Pi^{(2)\bm{k}}_{\mu\sigma}\tl\Pi^{(2)\bm{k}}_{\nu\rho}-\frac{2}{d-1}\tl\Pi^{(2)\bm{k}}_{\mu\nu}\tl\Pi^{(2)\bm{k}}_{\rho\sigma}\right)
\ee 
where we use these modified projectors in \equref{eq: new differential operators}.

We finally note the tensor structure of vertex factors given in \equref{eq: modified vertex factors}:
\be
\tl V^{\mu\nu\rho}_{\bm{v}_1,\bm{v}_2,\bm{v}_3}\equiv{}&
\eta^{\mu\nu}\left(\pdr{}{(\bm{v}_1)_\rho}-\pdr{}{(\bm{v}_2)_\rho}\right)+\text{ permutations}
\\
\tl V^{\mu\nu\rho\sigma\lambda\kappa}_{\bm{v}_1, \bm{v}_2, \bm{v}_3}\equiv{}&{} 
\left(
\eta^{\rho\lambda}\eta^{\sigma\kappa}\frac{\partial^2}{\partial(\bm{v}_2)_\mu \partial(\bm{v}_3)_\nu }
-2\eta^{\nu\lambda}\eta^{\sigma\kappa}\frac{\partial^2}{\partial(\bm{v}_2)_\mu \partial(\bm{v}_3)_\rho}
\right)+\text{ permutations}
\ee
with which one can define the full modified differential operator $\tl\cD$:
\be 
\label{eq: definition of d tilde}
\tl\cD^{m,n,r,t}_\text{gluon}\equiv& \left(
\prod\limits_{c=1,4,7,\dots}^{1+3(r-1)}
\tl V^{\rho_c\rho_{c+1}\rho_{c+2}}_{\bm{v}'_c,\bm{v}'_{c+1},\bm{v}'_{c+2}}
\right)
\left(
\prod\limits_{e=1,5,9,\dots}^{1+4(t-1)}
V^{\r_{e}\r_{e+1}\r_{e+2}\r_{e+3}}
\right)
\left(
\prod\limits_{a=1}^{m}
\bm{\epsilon}^a_{\mu_a}
\right)
\left(
\prod\limits_{b=1}^{n}
\tl\cD_{\nu_{2b-1}\nu_{2b}}^{(a_b)\bm{v}_b}
\right)
\\
\tl\cD^{m,n,r}_\text{graviton}\equiv& \left(
\prod\limits_{c=1,4,7,\dots}^{1+3(r-1)}
\tl V^{\rho_{2c-1}\rho_{2c}\rho_{2c+1}\rho_{2c+2}\rho_{2c+3}\rho_{2c+4}}_{\bm{v}'_1, \bm{v}'_2, \bm{v}'_3}
\right)
\left(
\prod\limits_{a=1}^{m}
\bm{\epsilon}^a_{\mu_{2a-1}\mu_{2a}}
\right)
\left(
\prod\limits_{b=1}^{n}
\tl\cD_{\nu_{4b-3}\nu_{4b-2}\nu_{4b-1}\nu_{4b}}^{(a_b)\bm{v}_b}
\right)
\ee 
with which we write down the Witten diagrams in terms of the scalar factors in \equref{eq: diffrential form of loop level Witten diagrams}.

\subsection{On integration of products of Bessel-type functions}
\label{sec: generic three point interactions}

We know  in momentum space formalism that the bulk point integrals we need to compute take the form
\be 
\int\limits_0^\infty z^{\lambda-1}E_\mu(a z)E_\nu(b z)E_\rho(c z)dz
\ee 
for three point interactions, where $E_a(x)\in\{J_a(x),K_a(x)\}$. In \cite{10.1093/qmath/os-6.1.52} Rice uses contour manipulations to compute such integrals in terms of Appell's hypergeometric function if $E=J$, for which the result reads as
\begin{multline}
\label{eq: JJJ in arbitrary dimensions}
\int\limits_0^\infty z^{\lambda-1}J_\mu(a z)J_\nu(b z)J_\rho(c z)dz=\frac{2^{\lambda-1}a^{\mu } b^{\nu }  \Gamma \left(\frac{\lambda +\mu +\nu +\rho}{2}\right )}{c^{ \lambda + \mu +\nu }\Gamma (\mu +1) \Gamma (\nu +1) \Gamma \left(1-\frac{\lambda +\mu +\nu -\rho}{2}\right )}
\\\x F_4\left(\frac{\lambda +\mu +\nu -\rho}{2}
,\frac{\lambda +\mu +\nu +\rho}{2};\mu +1,\nu +1;\frac{a^2}{c^2},\frac{b^2}{c^2}\right)\\\qquad\text{ for }\Re\left(\lambda+\mu+\nu+\rho\right)>0\;,\quad\Re\left(\lambda\right)<\frac{5}{2}\;,\quad c>a+b
\end{multline}
Same result has been computed independently by Bailey in \cite{doi:10.1112/plms/s2-40.1.37} who first uses hypergeometric identities to derive
\begin{multline}
\label{eq: JJK in arbitrary dimensions}
\int\limits_0^\infty z^{\lambda-1}J_\mu(a z)J_\nu(b z)K_\rho(c z)dz=\frac{2^{\lambda-2}a^{\mu } b^{\nu }  \Gamma \left(\frac{\lambda +\mu +\nu +\rho}{2}\right )\Gamma \left(\frac{\lambda +\mu +\nu -\rho}{2}\right )}{c^{ \lambda + \mu +\nu }\Gamma (\mu +1) \Gamma (\nu +1)}
\\\x F_4\left(\frac{\lambda +\mu +\nu -\rho}{2}
,\frac{\lambda +\mu +\nu +\rho}{2};\mu +1,\nu +1;-\frac{a^2}{c^2},-\frac{b^2}{c^2}\right)\\\qquad\text{ for }\Re\left(\lambda+\mu+\nu\right)>\abs{\Re\left(\rho\right)}\;,\quad\Re\left(c\pm ia\pm ib\right)>0
\end{multline} 
and then uses analytic continuation from BesselJ to BesselK to get \equref{eq: JJJ in arbitrary dimensions}. The identity he uses is
\be 
i\pi J_\mu(z)=e^{-i\pi\mu/2}K_\mu(-iz)-e^{i\pi\mu/2}K_\mu(iz)\quad\forall z>0
\ee 
and he argues that the transition is valid as the the integrand still converges. As $z^aK_a(z)$ better converges for $z\rightarrow\infty$ and is still convergent for $z\rightarrow0$, we can replace  $z^aJ_a(z)$ with $z^aK_a(z)$ where we can use the identity
\be 
K_{\mu }(z)=\frac{1}{2} \pi  \csc (\pi  \mu ) \left(e^{i\pi\mu/2}
J_{-\mu }(i z)-e^{-i\pi\mu/2} J_{\mu }(i z)\right)\quad \forall z>0
\ee
which means
\begin{multline}
\int\limits_0^\infty z^{\lambda-1}J_\mu(a z)K_\nu(b z)K_\rho(c z)dz=\Bigg[\frac{\Gamma (\nu )\Gamma \left(\frac{\lambda +\mu -\nu -\rho
	}{2}\right) \Gamma \left(\frac{\lambda +\mu -\nu +\rho }{2}\right)}{2^{3-\lambda } c^{\lambda } \left(\frac{c}{a}\right)^{\mu }
	\left(\frac{b}{c}\right)^{\nu }\Gamma (\mu +1)}
\\\x F_4\left(\frac{\lambda +\mu -\nu -\rho}{2}
,\frac{\lambda +\mu -\nu +\rho}{2};1+\mu ,1-\nu;-\frac{a^2}{c^2},\frac{b^2}{c^2}\right)\Bigg]
+ \left(\nu \rightarrow -\nu\right)
\\\qquad\text{ for }\Re\left(\lambda+\mu\pm\nu\right)>\abs{\Re\left(\rho\right)}\;,\quad c>b>0\;,\quad a>0
\end{multline}

\subsection{Computing loop integrals via standard QFT tricks}
\label{sec: loop integrals}
In this appendix we will review the solution of loop integrals via Feynman parametrization, a standard trick known from QFT. The general form of integrals of interest are
\be 
\cI=\int\limits_{\mathbb{R}^{d-1,1}} d^dl\frac{\left(\bm{l}\.\bm{v}_1\right)\cdots \left(\bm{l}\.\bm{v}_m\right)}{(a_1+(\bm{b}_1+\bm{l})^2)\dots (a_n+(\bm{b}_n+\bm{l})^2)}l^{2j}
\ee 
which can be parameterized with the Feynman trick as
\be 
\cI
=(n-1)!\int_0^1du_1\dots du_{n-1}\int\limits_{\mathbb{R}^{d-1,1}} d^dl\frac{\prod\limits_{a=1}^m\left(\bm{l}\.\bm{v}_a\right)l^{2j}}{\left[\sum\limits_{k=1}^nu_k(a_k+(\bm{b}_k+\bm{l})^2)\right]^n}
\ee 
for 
\be 
u_n\equiv1 -\sum\limits_{i=1}^{n-1}u_i
\ee 
We can then use
\be 
\sum\limits_{k=1}^nu_k(a_k+(\bm{b}_k+\bm{l})^2)=\left(\sum\limits_{i=1}^nu_i\right)\left[\left(\bm{l}+\frac{\sum u_i \bm{b}_i}{\sum u_i}\right)^2+\frac{\sum u_i(a_i+b_i^2)}{\left(\sum u_i\right)}-\frac{\left(\sum u_i\bm{b}_i\right)^2}{\left(\sum u_i\right)^2}\right]
\ee 
and shift the integration parameter to obtain
\begin{multline}
\cI
=(n-1)!\int_0^1du_1\dots du_{n-1}
\int\limits_{\mathbb{R}^{d-1,1}} d^dl\frac{\prod\limits_{a=1}^m\left(\bm{l}\.\bm{v}_a-\sum\limits_{i=1}^n u_i \bm{b}_i\.\bm{v}_a\right)\left(\bm{l}-\sum\limits_{i=1}^n u_i \bm{b}_i\right)^{2j}}{\left[
	l^2+\left(\sum\limits_{i=1}^n u_i(a_i+b_i^2)\right)-\left(\sum\limits_{i=1}^n u_i\bm{b}_i\right)^2	
	\right]^n}
\end{multline}
which we can rewrite as 
\begin{multline}
\label{eq: generic momentum integral middle step}
\cI
=
\sum\limits_{\alpha=0}^j\sum\limits_{\beta=0}^{j-\alpha}
\sum\limits_{\substack{i_1=0,1\\i_2=0,1\\\dots\\i_m=0,1}}
(n-1)!
\int_0^1du_1\dots du_{n-1}
c_{i_1\dots i_m}^{\alpha,\beta}
\int\limits_{\mathbb{R}^{d-1,1}} d^dl\frac{\prod\limits_{a=1}^m\left(\bm{l}\.\bm{v}_a\right)^{i_a}\left(\bm{l}\.\sum\limits_{i=1}^n u_i \bm{b}_i\right)^\alpha l^{2\beta}}{\left[
	l^2+\left(\sum\limits_{i=1}^n u_i(a_i+b_i^2)\right)-\left(\sum\limits_{i=1}^n u_i\bm{b}_i\right)^2	
	\right]^n}
\end{multline}
for
\be 
c_{i_1\dots i_m}^{\alpha,\beta}\equiv (-2)^\alpha\binom{j}{\alpha}\binom{j-\alpha}{\beta}\left(\sum\limits_{i=1}^n u_i \bm{b}_i\right)^{2j-2\alpha-2\beta}\prod\limits_{a=1}^m\left(-\sum\limits_{i=1}^n u_i \bm{b}_i\.\bm{v}_a\right)^{1-i_a}
\ee 

We note that the integrand is  a function of $l^2$ only except for $\left(\bm{l}\.\bm{v}_a\right)^{i_a}\left(\bm{l}\.\sum\limits_{i=1}^n u_i \bm{b}_i\right)^\alpha$ where the exponents are integers, hence the Lorentz symmetry allows us to make the replacements
\be
\bm{l}_{\mu_1}\bm{l}_{\mu_2}{\dots} \bm{l}_{\mu_{2n+1}}
\quad&\rightarrow\quad0\\
\bm{l}_{\mu_1}\bm{l}_{\mu_2}{\dots}\bm{l}_{\mu_{2n}}\quad&\rightarrow\quad\frac{l^{2n}}{\prod\limits_{k=1}^n\left(d-1+(2k-1)!!\right)}\sum\limits_{p\in P_{2n}^2}\prod\limits_{\{a,b\}\in p}\eta_{ab}
\ee
where the sum is over all distinct ways of partitioning $\{1,2,\dots, 2n\}$ into pairs $\{a,b\}$, and the product is over the pairs contained in $p$. For example,
\be 
\bm{l}_\mu\bm{l}_\nu\quad&\rightarrow\quad\frac{l^2}{d}\eta_{\mu\nu}\\
\bm{l}_\mu\bm{l}_\nu\bm{l}_\rho\bm{l}_\lambda\quad&\rightarrow\quad\frac{l^4}{d(d+2)}\left(\eta_{\mu\nu}\eta_{\rho\lambda}+\eta_{\mu\rho}\eta_{\nu\lambda}+\eta_{\mu\lambda}\eta_{\nu\rho}\right)
\ee 

We can now reexpress \equref{eq: generic momentum integral middle step} as
\begin{multline}
\label{eq: middle step for cI}
\cI
=
\sum\limits_{\alpha=0}^j\sum\limits_{\beta=0}^{j-\alpha}
\sum\limits_{\substack{i_1=0,1\\i_2=0,1\\\dots\\i_m=0,1}}
(n-1)!
\int_0^1du_1\dots du_{n-1}
c_{i_1\dots i_m}^{\alpha,\beta}
\frac{\sum\limits_{p\in \mathcal{P}_{i_1\dots i_m}^\alpha}\prod\limits_{\{\bm{x},\bm{y}\}\in p}(\bm{x}\.\bm{y})}{\prod\limits_{k=1}^{\left(\alpha+\sum_a i_a\right)/2}\left(d-1+(2k-1)!!\right)}
\\\x
\int\limits_{\mathbb{R}^{d-1,1}} d^dl\frac{l^{\alpha+2\beta+\sum_a i_a}}{\left[
	l^2+\left(\sum\limits_{i=1}^n u_i(a_i+b_i^2)\right)-\left(\sum\limits_{i=1}^n u_i\bm{b}_i\right)^2	
	\right]^n}
\end{multline}
where $\mathcal{P}_{i_1\dots i_m}^\alpha$ is the list which has the element $\bm{v}_a$ $i_a$ times, and the element $\sum\limits_{i=1}^n u_i \bm{b}_i$ $\alpha$ times; for example
\be 
\label{eq: example set partitions}
\mathcal{P}_{1,1}^1=\left\{\bm{v}_1,\bm{v}_2,\sum\limits_{i=1}^n u_i \bm{b}_i\right\}\;,\quad 
\mathcal{P}_{1,0,1}^2=\left\{\bm{v}_1,\bm{v}_3,\sum\limits_{i=1}^n u_i \bm{b}_i,\sum\limits_{i=1}^n u_i \bm{b}_i\right\}
\ee 
Note that the partitioning of $p\in \mathcal{P}_{i_1\dots i_m}^\alpha$ is only possible if  $\mathcal{P}$ has even number of elements, hence
\be 
\sum\limits_{p\in \mathcal{P}_{1,1}^1}\prod\limits_{\{\bm{x},\bm{y}\}\in p}(\bm{x}\.\bm{y})=0
\ee 
whereas 
\be 
\sum\limits_{p\in \mathcal{P}_{1,0,1}^2}\prod\limits_{\{\bm{x},\bm{y}\}\in p}(\bm{x}\.\bm{y})=\bm{v}_1\.\bm{v}_3\left(\sum\limits_{i=1}^n u_i \bm{b}_i\right)\.\left(\sum\limits_{i=1}^n u_i \bm{b}_i\right)+2\bm{v}_1\.\left(\sum\limits_{i=1}^n u_i \bm{b}_i\right)\bm{v}_3\.\left(\sum\limits_{i=1}^n u_i \bm{b}_i\right)
\ee 
This is just the realization of the fact that integration volume is invariant under  $\bm{l}\rightarrow -\bm{l}$, hence integrands with odd number of $\bm{l}$ vanish.

We are now left with the $l-$integration in \equref{eq: middle step for cI}. To proceed, we first use the well-known identity
\be 
\int\limits_{\R^{d-1,1}} \frac{d^dl}{(2\pi)^d}\frac{1}{\left[l^2-\De\right]^n}=\frac{i(-1)^n}{(4\pi)^{d/2}}\frac{\Gamma(n-d/2)}{\Gamma(n)}\De^{d/2-n}
\ee 
which can be generalized as 
\be 
\int\limits_{\mathbb{R}^{d-1,1}} \frac{d^dl}{(2\pi)^d}\frac{l^{2m}}{\left[l^2-\De\right]^n}=&\sum _{k=0}^m  \Delta ^{-k+m} \binom{m}{k}\int\limits_{\mathbb{R}^{d-1,1}} \frac{d^dl}{(2\pi)^d}\frac{1}{\left[l^2-\De\right]^{n-k}}
\\=&\frac{i 2^{-d} \pi ^{-d/2} (-1)^n \Gamma \left(n-\frac{d}{2}\right) }{\Gamma (n)}\pFq{2}{1}{-m,1-n}{\frac{d-2 n+2}{2}}{1}\Delta
^{\frac{d}{2}+m-n}
\ee 

We can now write down the final result:
\begin{multline}
\label{eq: generic volume integral}
\int\limits_{\mathbb{R}^{d-1,1}} d^dl\frac{\left(\bm{l}\.\bm{v}_1\right)\cdots \left(\bm{l}\.\bm{v}_m\right)}{(a_1+(\bm{b}_1+\bm{l})^2)\dots (a_n+(\bm{b}_n+\bm{l})^2)}l^{2j}
=
i \pi ^{d/2} (-1)^n \Gamma \left(n-\frac{d}{2}\right)
\\\x 
\sum\limits_{\alpha=0}^j\sum\limits_{\beta=0}^{j-\alpha}
\sum\limits_{\substack{i_1=0,1\\i_2=0,1\\\dots\\i_m=0,1}}
\frac{\pFq{2}{1}{-\sigma,1-n}{\frac{d-2 n+2}{2}}{1}}{\prod\limits_{k=1}^{\sigma-\b}\left(d-1+(2k-1)!!\right)}
\int_0^1du_1\dots du_{n-1}
\left(\sum\limits_{p\in \mathcal{P}_{i_1\dots i_m}^\alpha}\prod\limits_{\{\bm{x},\bm{y}\}\in p}(\bm{x}\.\bm{y})\right)
c_{i_1\dots i_m}^{\alpha,\beta}
\Delta^{\frac{d}{2}+\sigma-n}
\end{multline}
for
\be 
c_{i_1\dots i_m}^{\alpha,\beta}\equiv& (-2)^\alpha\binom{j}{\alpha}\binom{j-\alpha}{\beta}\left(\sum\limits_{i=1}^n u_i \bm{b}_i\right)^{2j-2\alpha-2\beta}\prod\limits_{a=1}^m\left(-\sum\limits_{i=1}^n u_i \bm{b}_i\.\bm{v}_a\right)^{1-i_a}
\\
\sigma \equiv&\frac{1}{2}\left(\a +\sum\limits_{a=1}^m i_a\right)+\b
\\
\Delta \equiv&\left(\sum\limits_{i=1}^n u_i\bm{b}_i\right)^2 - \left(\sum\limits_{i=1}^n u_i(a_i+b_i^2)\right)
\\
u_n\equiv& 1 -\sum\limits_{i=1}^{n-1}u_i
\ee 
where the set $\mathcal{P}_{i_1\dots i_m}^\alpha$ is defined and detailed around \equref{eq: example set partitions}.

As an example, we see that
\begin{multline}
\int\limits_{\mathbb{R}^{d-1,1}} d^d\bm{l}
\frac{(\bm{l}\.\bm{v}_2)^2(\bm{l}\.\bm{v}_3)\left(\bm{l}\.\bm{v}_4\right)
}{\abs{\bm{k}+\bm{l}}^2l^2}=
i \pi ^{d/2} \Gamma \left(2-\frac{d}{2}\right)
\sum\limits_{\substack{i_1=0,1\\i_2=0,1\\i_3=0,1\\i_4=0\text{ or }1 \text{ such that }\\i_1+i_2+i_3+i_4\in 2\mathbb{N}}}
\frac{\pFq{2}{1}{-\sigma,-1}{\frac{d-2}{2}}{1}}{\prod\limits_{k=1}^{\sigma}\left(d-1+(2k-1)!!\right)}
\\\x
\left(\bm{k}\.\bm{v}_2\right)^{1-i_1}\prod_{j=2}^4\left(\bm{k}\.\bm{v}_j\right)^{1-i_j}
\left(\bm{k}\.\bm{k} \right)^{\frac{d}{2}+\sigma-2}
\left(\sum\limits_{p\in \mathcal{P}_{i_1i_2i_3i_4}^0}\prod\limits_{\{\bm{x},\bm{y}\}\in p}(\bm{x}\.\bm{y})\right)
\int_0^1du
u^{\frac{d}{2}+2-\sigma}
(u-1)^{\frac{d}{2}-2+\sigma}
\end{multline}
for 
\be 
\sigma=(i_1+i_2+i_3+i_4)/2
\ee 
which then becomes
\begin{multline}
\label{eq: volume integration example}
\int\limits_{\mathbb{R}^{d-1,1}} d^d\bm{l}
\frac{(\bm{l}\.\bm{v}_2)^2(\bm{l}\.\bm{v}_3)\left(\bm{l}\.\bm{v}_4\right)
}{\abs{\bm{k}+\bm{l}}^2l^2}
= \frac{i^{d-1} \pi ^{\frac{d}{2}+1} \csc \left(\frac{\pi  d}{2}\right) \Gamma
	\left(\frac{d}{2}+3\right)}{\Gamma (d+2)}
\Bigg(
k^{d-4}\left(\bm{k}\.\bm{v}_2\right)^{2}\left(\bm{k}\.\bm{v}_3\right)\left(\bm{k}\.\bm{v}_4\right)
\\
-\frac{d k^{d-2}}{d+4}\Big[\left(\bm{k}\.\bm{v}_2\right)^2\left(\bm{v}_3\.\bm{v}_4\right)
+
2\left(\bm{k}\.\bm{v}_2\right)\left(\bm{k}\.\bm{v}_3\right)\left(\bm{v}_2\.\bm{v}_4\right)
+2
\left(\bm{k}\.\bm{v}_2\right)\left(\bm{k}\.\bm{v}_4\right)\left(\bm{v}_2\.\bm{v}_3\right)
\\+
\left(\bm{k}\.\bm{v}_3\right)\left(\bm{k}\.\bm{v}_4\right)\left(\bm{v}_2\.\bm{v}_2\right)
\Big]
+\frac{d k^d}{d+4}\Big[\left(\bm{v}_2\.\bm{v}_2\right)\left(\bm{v}_3\.\bm{v}_4\right)+2\left(\bm{v}_2\.\bm{v}_3\right)\left(\bm{v}_2\.\bm{v}_4\right)\Big]
\Bigg)
\end{multline}

% END
\bibliographystyle{utphys}
\bibliography{references}{}
\end{document}